\documentclass[aip,jap,reprint,superscriptaddress]{revtex4-2}

\usepackage[bookmarksnumbered,hyperfootnotes=false]{hyperref} % Hyperlinks
\usepackage{graphicx} % Grafiken
\usepackage{textcomp}
\usepackage{dcolumn}
\usepackage{bm}
\usepackage{amssymb}

\usepackage[utf8]{inputenc}
\usepackage[T1]{fontenc}
\usepackage{mathptmx}
\usepackage{color}

\definecolor{purple}{cmyk}{.47,1,0,.03}
\definecolor{orange}{cmyk}{0,.51,1,.0}
\definecolor{brown}{cmyk}{0,1,1,.47}

\begin{document}

%\preprint{}

%\preprint{AIP/123-QED}

\title{Impact of severe plastic deformation on the relaxation of glassy and supercooled liquid states of amorphous Pd$_{40}$Ni$_{40}$P$_{20}$}


\author{Afrouz Hassanpour}
%\thanks{}
\affiliation{Institut f\"ur Materialphysik, Westf\"alische Wilhelms-Universit\"at M\"unster, Wilhelm-Klemm-Str. 10, 48149 M\"unster, Germany}

\author{Sven Hilke}
\email[Corresponding author: ]{sven.hilke@uni-muenster.de}
\affiliation{Institut f\"ur Materialphysik, Westf\"alische Wilhelms-Universit\"at M\"unster, Wilhelm-Klemm-Str. 10, 48149 M\"unster, Germany}

\author{Harald R\"osner}
\email[Corresponding author: ]{rosner@uni-muenster.de}
\affiliation{Institut f\"ur Materialphysik, Westf\"alische Wilhelms-Universit\"at M\"unster, Wilhelm-Klemm-Str. 10, 48149 M\"unster, Germany}

\author{Sergiy V. Divinski}
\affiliation{Institut f\"ur Materialphysik, Westf\"alische Wilhelms-Universit\"at M\"unster, Wilhelm-Klemm-Str. 10, 48149 M\"unster, Germany}

\author{Gerhard Wilde}
\email[Corresponding author: ]{gwilde@uni-muenster.de}
\affiliation{Institut f\"ur Materialphysik, Westf\"alische Wilhelms-Universit\"at M\"unster, Wilhelm-Klemm-Str. 10, 48149 M\"unster, Germany}


\date{\today}

\begin{abstract}
The impact of severe plastic deformation by high-pressure torsion on the relaxation of the glassy and supercooled liquid states of Pd$_{40}$Ni$_{40}$P$_{20}$ was investigated using a combination of differential scanning calorimetry, low-temperature heat capacity and fluctuation electron microscopy. The changes in the calorimetric signals due to deformation and subsequent heat treatments were analyzed and a correlation between deformation (rejuvenation) and annealing (relaxation) was found in relation to medium-range order (MRO). Moreover, a coupling between the occurrence of an exothermic peak in the supercooled liquid state and specific changes in the MRO types were identified. These findings are comprehended in a potential energy landscape scheme offering a new approach for MRO engineering of glasses.

\end{abstract}


\maketitle


%%%%%%%%%%%%%%%%%%%%%%%%%%%%%%%%%%%%%%%%%%%%%%%%%%%%%%%%%%%%%%%%%%%%%%%%%%%%
%%%%%%%%%%%%%%%%%%%%%%%%%%%%%%%%%%%%%%%%%%%%%%%%%%%%%%%%%%%%%%%%%%%%%%%%%%%%
%%%%%%%%%%%%%%%%%%%%%%%%%%%%%%%%%%%%%%%%%%%%%%%%%%%%%%%%%%%%%%%%%%%%%%%%%%%%
%%%%%%%%%%%%%%%%%%%%%%%%%%%%%%%%%%%%%%%%%%%%%%%%%%%%%%%%%%%%%%%%%%%%%%%%%%%%
%%%%%%%%%%%%%%%%%%%%%%%%%%%%%%%%%%%%%%%%%%%%%%%%%%%%%%%%%%%%%%%%%%%%%%%%%%%%
\section{Introduction}
Monolithic metallic glasses are formed from the liquid phase upon cooling when crystallization is avoided. Most metallic liquids have low viscosity favoring nucleation and growth of crystalline seeds which makes glass formation difficult or even unlikely to happen. Eutectic alloys, such as Pd$_{40}$Ni$_{40}$P$_{20}$ having a deep melting point, are predestined to produce bulk metallic glasses (BMGs) since slower cooling rates ($\sim10^1$\,K/s) lead to the desired glassy state \cite{mitsch2000heat,mitrofanov2015impact,wilde1996thermophysical,wilde1997stability}. Moreover, Pd$_{40}$Ni$_{40}$P$_{20}$ exhibits good glass-forming ability \cite{he1996bulk,wilde1996thermophysical,wilde1997stability} and mechanical properties (ductility) \cite{nollmann2016impact,Davani2019,hilke2020role}. Depending on the cooling rate, the cast-material inherently attains variations of local atomic structures corresponding to higher and lower energy states \cite{stillinger1995topographic}. The higher energy states of the as-cast material can be manipulated by annealing which causes relaxation, i.e. aging, of the glassy state to lower energies \cite{moynihan1976ha,van1983kinetics,mitsch2000heat}. The opposite process is called rejuvenation, that is bringing the glassy material to higher energy states. This can be achieved for instance by deformation \cite{zhou2019two,hubek2018impact}, cycling in the elastic regime \cite{ross2017linking} or cryogenic cycling \cite{ketov2015rejuvenation,guo2018rejuvenation,bian2017cryogenic} which locally ''warps`` the potential energy landscape (PEL). However, the underlying mechanisms for aging and rejuvenation are still not fully understood. 

In this study we address relaxation and rejuvenation of Pd$_{40}$Ni$_{40}$P$_{20}$ by annealing both severely plastically deformed and as-cast material below the glass transition temperature. Calorimetric measurements of severely deformed material revealed a significant exothermic signal in the supercooled liquid below the crystallization peak. To unveil the origin of this exothermic peak its thermodynamic state was frozen-in by rapid-quenching in the calorimeter. The corresponding materials' states were characterized with respect to their medium-range order (MRO) to show the structural changes taking place during aging and rejuvenation. It worth noting that the individual materials' states investigated were fully amorphous after all treatments. The obtained results identify a coupling between the MRO with the thermo-mechanical history of the BMG and the corresponding PEL.

The current paper is structured as follows: First, we show that thermo-mechanical treatments via a combination of room temperature high pressure torsion (HPT) and subsequent annealing give rise to significant changes in the calorimetric response in the supercooled liquid region measured by differential scanning calorimetry (DSC). Low-temperature heat capacity measurements (PPMS) were performed in order to reveal changes in the atomic vibrational spectra which are related to the excess free volume distribution \cite{turnbull1961free,grest1981liquids} and thus atomic arrangements. To investigate the corresponding  amorphous structure in detail, fluctuation electron microscopy (FEM) was employed to determine the MRO fingerprint (size \cite{hwang2011variable,Davani2019}, relative volume fraction \cite{bogle2007quantifying,bogle2010size} and types \cite{hilke2019influence,pekin2019direct}) of each materials' state \cite{voyles2002fluctuation,voyles2002fluctuatione,voyles2000fluctuation,voyles2003medium,hwang2011variable,treacy1996variable,treacy2005fluctuation,iwai1999method}. Finally, these findings are discussed in terms of a PEL scheme offering a new approach for MRO engineering of glasses.

%%%%%%%%%%%%%%%%%%%%%%%%%%%%%%%%%%%%%%%%%%%%%%%%%%%%%%%%%%%%%%%%%%%%%%%%%%%%
%%%%%%%%%%%%%%%%%%%%%%%%%%%%%%%%%%%%%%%%%%%%%%%%%%%%%%%%%%%%%%%%%%%%%%%%%%%%
%%%%%%%%%%%%%%%%%%%%%%%%%%%%%%%%%%%%%%%%%%%%%%%%%%%%%%%%%%%%%%%%%%%%%%%%%%%%
%%%%%%%%%%%%%%%%%%%%%%%%%%%%%%%%%%%%%%%%%%%%%%%%%%%%%%%%%%%%%%%%%%%%%%%%%%%%
%%%%%%%%%%%%%%%%%%%%%%%%%%%%%%%%%%%%%%%%%%%%%%%%%%%%%%%%%%%%%%%%%%%%%%%%%%%%

\section{Materials and Methods}

Pd$_{40}$Ni$_{40}$P$_{20}$ BMG was produced by ingot copper mold casting under Ar atmosphere. Prior to casting, the sample was fluxed with boron oxide (B$_2$O$_3$) and subsequently cycled three times between 900 and 1000\,\textdegree C for purification \cite{wilde2006bulk}. The sample size of the ''as-cast`` state was $30\times10\times1$\,mm$^3$.

\begin{table*}[htbp]
	\centering
	\caption{Compilation of samples and analytical techniques.}
	\label{tab:TAB1}
	%\resizebox{\textwidth}{!}{%
	\begin{ruledtabular}
	\begin{tabular}{lc|cccc}
		& ''sample state`` (\textbf{color code})  & XRD & DSC {[}Fig.~\ref{fig:FIG1}{]} & PPMS {[}Fig.~\ref{fig:FIG2}{]} & FEM {[}Figs.~\ref{fig:FIG3} and ~\ref{fig:FIG6}{]} \\ \hline \hline
		1. & ''as-cast`` (\textbf{black})                                                                                       & $\checkmark$                                        & $\checkmark$                                               & $\checkmark$                                                & $\checkmark$                                                        \\ \hline
		2. &\color{blue}{''10 HPT`` (\textbf{blue})}                                                                                    & $\checkmark$                                               & $\checkmark$                                               & $\checkmark$                                                & $\checkmark$                                                         \\ \hline \hline
		3. & \color{red}{''as-cast + annealed 561\,K 2\,h`` (\textbf{red})}                                & $\checkmark$                                               & $\checkmark$                                               & $\checkmark$                                                & $\checkmark$                                                         \\ \hline
		4. & \color{purple}{''10 HPT + annealed 561\,K 2\,h`` (\textbf{purple})}                               & $\checkmark$                                               & $\checkmark$                                               & $\checkmark$                                                & $\checkmark$                                                       \\ \hline \hline
		5. & \color{orange}{''10 HPT + heated to 648\,K`` (\textbf{orange})}                                                    & $\checkmark$                                               & $-$                                               & $-$                                                & $\checkmark$                                                        \\ \hline
		6. & \color{brown}{''10 HPT + annealed 561\,K 2\,h + heated to 648\,K`` (\textbf{brown})} & $\checkmark$                                               & $-$                                               & $-$                                                & $\checkmark$                                                        \\ 
	\end{tabular}%
	%}
	\end{ruledtabular}
\end{table*}

%%%%%%%%%%%%%%%%%%%%%%%%%%%%%%%%%%%%%%%%%%%%%%%%%%%%%%%%%%%%%%%%%%%%%%%%%%%%

HPT deformation was performed applying $ N = 10 $ rotations. Disk-shaped samples were used having a diameter of 10\,mm and a thickness of 1\,mm. The samples were positioned between two anvils (one flat and one with a suitable cavity) and rotated with a speed of 1\,rpm under a quasi-hydrostatic pressure of 4\,GPa. The shear strain, $ \gamma $, estimated by $ \gamma = \frac{2\pi N}{h} \cdot \frac{r}{2}$, where $ r $ and $ h $ are the radius and thickness of the samples, at the half radius of the samples (positions, from where the samples were taken) was equal to $314\,\%$. Usually, the microstructure evolution across the disc diameter is heterogeneous. However, for $ N = 10 $ rotations the shear deformation can be treated as reasonably homogeneous except near the center and at the near-edge regions \cite{van2010effects,hobor2008high,zhilyaev2008using,wang2011atomic,boucharat2005nanocrystallization}. In the following, this sample state is called ''10 HPT``.

%%%%%%%%%%%%%%%%%%%%%%%%%%%%%%%%%%%%%%%%%%%%%%%%%%%%%%%%%%%%%%%%%%%%%%%%%%%%

The amorphous state of all samples was confirmed using a Siemens D5000 diffractometer with Cu-K$_\alpha$ radiation. DSC was performed on ''as-cast`` and ''10 HPT`` sample with a Perkin Elmer Diamond DSC device using a heating/cooling rate of 20\,K/min. All measurements shown are baseline corrected. Moreover, it is worth noting that the glass transition temperature T$_g$ is defined here as the temperature at which the integrated heat release corresponding to the supercooled liquid equals that of the relaxation state \cite{hubek2018impact, tool1931variations}. Relaxation experiments were performed under Ar atmosphere via isothermal holding in the DSC at 561\,K for 2 hours. These two additional sample states are called ''as-cast + annealed 561\,K 2\,h`` and ''10 HPT + annealed 561\,K 2\,h`` which were also investigated by DSC. Moreover, two more sample states were generated from the ''10 HPT`` and ''10 HPT + annealed 561\,K 2\,h`` sample states and ''frozen in`` by rapid quenching (100\,K/min) from the supercooled liquid at 648\,K in the DSC and called ''10 HPT + heated to 648\,K`` and ''10 HPT + annealed 561\,K 2\,h + heated to 648\,K``. A compilation of all samples and analytical techniques used in this study is given in Tab.~\ref{tab:TAB1}.

%%%%%%%%%%%%%%%%%%%%%%%%%%%%%%%%%%%%%%%%%%%%%%%%%%%%%%%%%%%%%%%%%%%%%%%%%%%%

Low-temperature heat capacity measurements were performed with a physical properties measurement system (PPMS, Quantum Design) in the temperature range between 1.9 and 100\,K \cite{schirmacher1998harmonic,bunz2014low,mitrofanov2015impact}. The specific heat capacity was measured at constant pressure. The samples were placed on a holder using low-temperature grease to give a good thermal contact. Each measurement consisted of two parts: (1) the so-called addenda, which determined the specific heat capacity of the holder and the grease; (2) under identical parameters, the specific heat capacity of the complete assembly, i.e. the holder, grease and sample. Subsequently, the specific heat capacity of the sample was obtained by subtracting the addenda from the second measurement \cite{hwang1997measurement}. Debye’s theory for the specific heat capacity of solids gives the phonon contribution to the total heat capacity of the metallic glass samples at low temperatures:
\begin{equation}
C_p^{Ph} = C_p - \gamma T \quad ,
\label{EQ:Cp}
\end{equation}
where $C_p^{Ph}$ is equal to $\beta T^3$ at low temperatures; $\gamma$ and $\beta$ are the electronic and phononic (Debye) contributions, respectively. $C_p$ denotes the measurements of the specific heat capacity as a function of temperature. The characteristic low-temperature anomaly of the specific heat capacity in glasses is termed ''boson excess peak`` \cite{kanazawa2001boson,li2006low}. Plotting $C_p^{Ph}/T^3$ versus $ T $ reveals the deviation from the Debye-like behavior at low temperatures due to scattering of the quasi-phonons of the glass at structural heterogeneities \cite{hubek2018impact,schirmacher1998harmonic,bunz2014low,mitrofanov2015impact,khonik2018boson}. 

%%%%%%%%%%%%%%%%%%%%%%%%%%%%%%%%%%%%%%%%%%%%%%%%%%%%%%%%%%%%%%%%%%%%%%%%%%%%
Samples, having a diameter of 3\,mm and a thickness of 70\,\textmu m, were prepared for transmission electron microscopy (TEM) by electropolishing. A TenuPol-5 device with a BK2 electrolyte \cite{kestel1986non} was used at a voltage of 16.8\,V keeping the temperature of the electrolyte between $-20$ and $-30$\,\textdegree C.   

%%%%%%%%%%%%%%%%%%%%%%%%%%%%%%%%%%%%%%%%%%%%%%%%%%%%%%%%%%%%%%%%%%%%%%%%%%%%

FEM is a microscopic technique which has been shown to be sensitive to higher order correlation functions such as g$_3$ and g$_4$ and hence to MRO in disordered materials \cite{voyles2002fluctuatione,voyles2000fluctuation,voyles2003medium,hwang2011variable,treacy1996variable,treacy2005fluctuation,iwai1999method,Davani2019,hilke2020role,hilke2019influence,davani2020shear}. A statistical analysis of the diffracted intensities $ I (\vec{k},R,\vec{r}) $ from nanometer-sized volumes was used by utilizing STEM microdiffraction \cite{voyles2002fluctuation}, where $\vec{k}$ is the scattering vector, $ R $ is the probe size and $\vec{r}$ is the position on the sample. Variable resolution FEM (VR-FEM) data sets were acquired by sampling with different probe sizes, $ R $, being sensitive to MRO at different length scales. The MRO was analyzed by calculating the normalized variance profiles using a pixel-by-pixel analysis according to the annular mean of variance image ($\Omega_{VImage} (k)$) \cite{daulton2010nanobeam}. A detailed description of the analysis method is given in Refs. \cite{Davani2019,hilke2020role} using the normalized variance:
\begin{equation}
V(|\vec{k}|,R) = \frac{\left\langle I^2 (\vec{k},R,\vec{r}) \right\rangle}{\left\langle I (\vec{k},R,\vec{r}) \right\rangle^2}-1 \quad.
\label{EQ:NVAR}
\end{equation}

\begin{figure*}[htbp]
	\includegraphics[width=\textwidth]{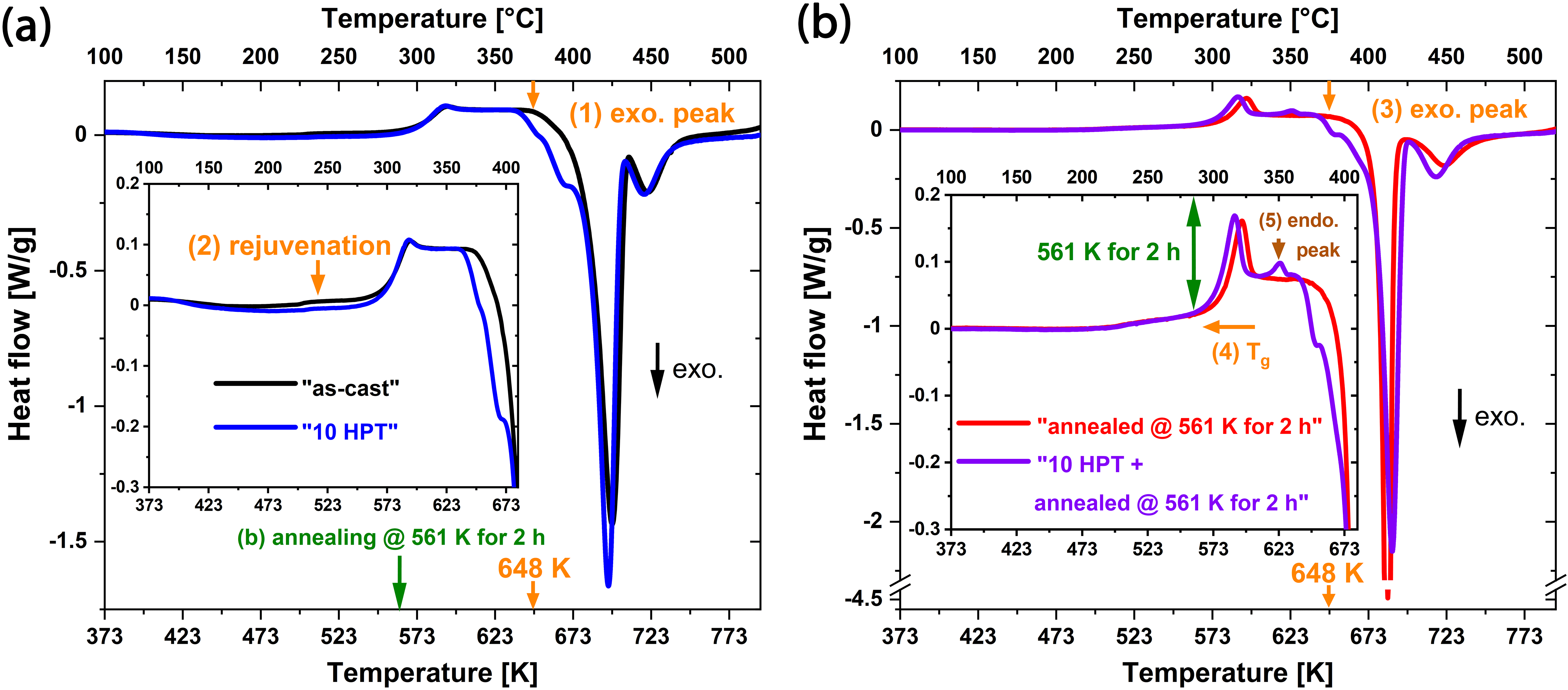}%
	\centering
	\caption{DSC charts of ''as-cast`` and ''10 HPT`` (a) as well as the corresponding annealed states (b) are shown.}
	\label{fig:FIG1}
\end{figure*}

VR-FEM experiments were performed using a Thermo Fisher Scientific FEI Themis G3 60-300 transmission electron microscope. The microscope was operated at an acceleration voltage of 300 kV with an extraction voltage of 3.45\,kV for the X-FEG. The nanobeam diffraction patterns (NBDPs) were acquired in \textmu P-STEM mode to assure parallel illumination using probe sizes between 0.8 and 4.2\,nm. Different probe sizes obtained by adjusting the semi-convergence angle \cite{yi2010flexible}. The corresponding foil thicknesses were calculated to be about $(0.6 \pm 0.12)$ mean free path (MFP) or $(45 \pm 9)$\,nm from individual electron energy loss (EEL) spectra using the log-ratio method \cite{malis1988eels}. The acquisition time was set to 3.5 seconds to ensure a good signal to noise ratio for the normalized variance analysis and to avoid Poisson noise \cite{yi2010flexible}. The acquisition of the NBDPs was performed with the full CCD camera at binning 4 with 512x512 pixels; the camera length was set to 60\,mm to access higher k-values. The probe diameters used for the acquisition of the NBDPs were measured with a Ceta camera prior to the experiments using Gatan Digital Micrograph and plugins by Mitchell \cite{mitchell2005scripting,mitchell2006determination}. It is emphasized that all experimental parameters, i.e. thickness, probe current (15\,pA), exposure time, were similar for all six samples. A minimum of 400 NBDPs were acquired for each sample and probe size. Thus, the corresponding (statistical) errors are small. X-Ray spikes on the CCD camera were removed from the individual NBDPs before analyzing the data further. The individual calculated normalized variance curves are shown in the Supplementary Material Fig. S1. The results of the FEM analyses are displayed in the form of Stratton-Voyles plots \cite{stratton2008phenomenological}, in which the peak intensity of the first normalized variance peak $V(k)$ at $k=4.8\,$nm$^{-1}$ is plotted against $1/R^2$\,\,\cite{stratton2007comparison}. Stratton-Voyles plots make an inspection of the individual treated materials' states with respect to the MRO changes feasible and easier to interpret. A more detailed and recent discussion on different models for describing FEM-based data can be found in Ref. \cite{Davani2019}. Moreover, in order to elucidate subtle differences in the amorphous structure between the individual materials' states differential curves $\Delta$V(k,R) of the normalized variances are analyzed.

%%%%%%%%%%%%%%%%%%%%%%%%%%%%%%%%%%%%%%%%%%%%%%%%%%%%%%%%%%%%%%%%%%%%%%%%%%%%
%%%%%%%%%%%%%%%%%%%%%%%%%%%%%%%%%%%%%%%%%%%%%%%%%%%%%%%%%%%%%%%%%%%%%%%%%%%%
%%%%%%%%%%%%%%%%%%%%%%%%%%%%%%%%%%%%%%%%%%%%%%%%%%%%%%%%%%%%%%%%%%%%%%%%%%%%
%%%%%%%%%%%%%%%%%%%%%%%%%%%%%%%%%%%%%%%%%%%%%%%%%%%%%%%%%%%%%%%%%%%%%%%%%%%%
%%%%%%%%%%%%%%%%%%%%%%%%%%%%%%%%%%%%%%%%%%%%%%%%%%%%%%%%%%%%%%%%%%%%%%%%%%%%

\section{Results}

\subsection{Differential Scanning Calorimetry}
Fig.~\ref{fig:FIG1}a shows the calorimetric results (DSC) for the ''as-cast`` and the ''10 HPT`` sample state. Plastic deformation (10 HPT) has two important effects on the calorimetric response of the sample: firstly, the appearance of an exothermic peak at 648\,K (1) at temperatures below the onset of the crystallization peak in the supercooled liquid region and secondly, rejuvenation (2), i.e. an increase of the relaxation peak occurring below the glass transition temperature (T$_g$). 

In comparison to Fig.~\ref{fig:FIG1}a, the effect of relaxation is shown in Fig.~\ref{fig:FIG1}b displayed by ''as-cast + annealed 561\,K 2\,h`` and ''10 HPT + annealed 561\,K 2\,h``. Annealing below the glass transition temperature T$_g$ (annealing for two hours at 561\,K) reduces the enthalpy of the exothermic peak in the supercooled region (3) from 1.34\,J/g (Fig.~\ref{fig:FIG1}a) to 0.49\,J/g (Fig.~\ref{fig:FIG1}b). A shift of about 6\,K in T$_g$ (4) is also noticeable. Moreover, an endothermic peak (5) occurs in the supercooled liquid at about 623\,K after annealing of the deformed sample. All observations are summarized in Tab.~\ref{tab:TAB2}.

\begin{figure}[htbp]
	\includegraphics[width=\columnwidth]{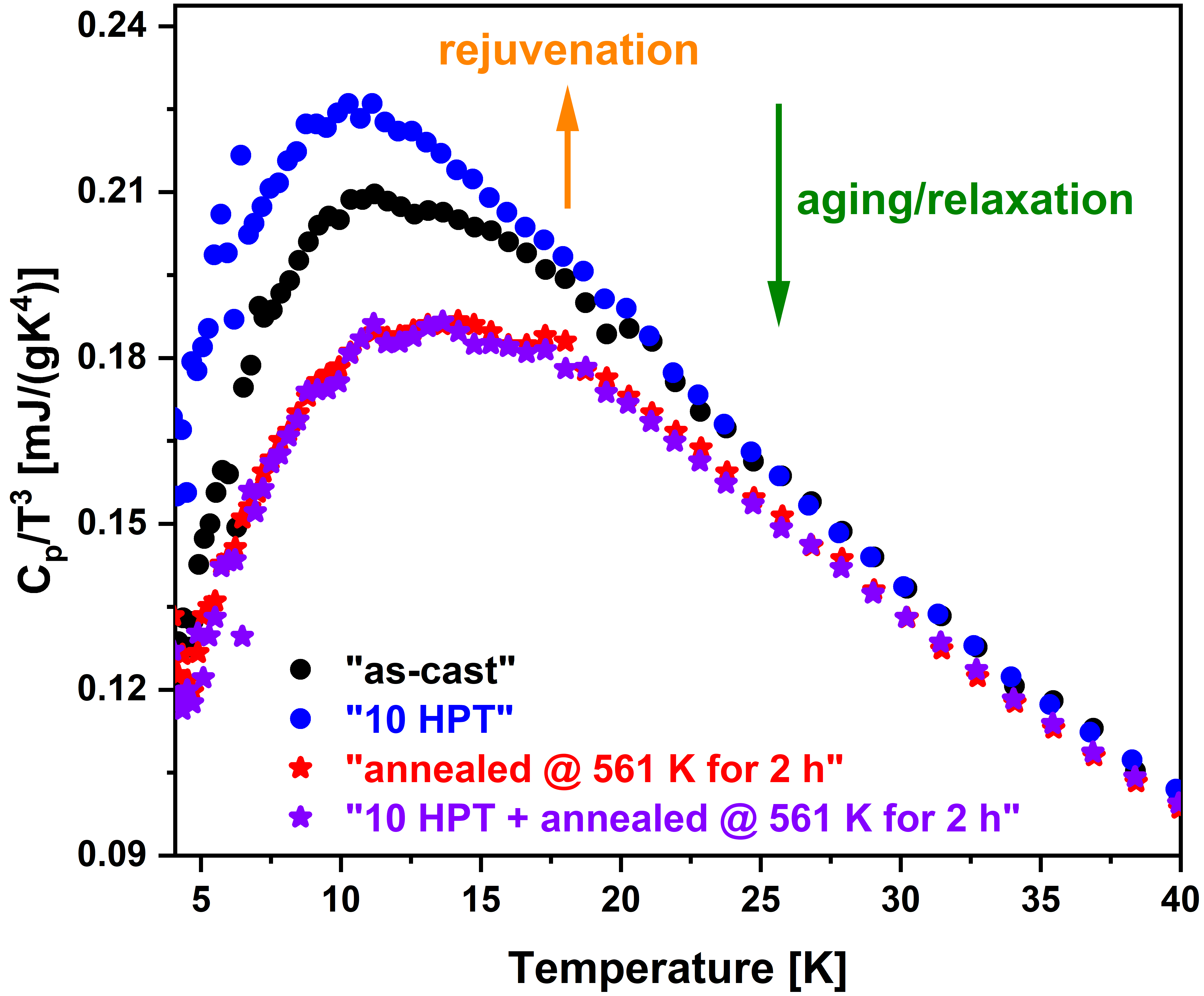}%
	\centering
	\caption{Boson peaks for as-cast, the 10 HPT sample as well as their subsequently annealed (at 561\,K for 2 hours) counterparts.}
	\label{fig:FIG2}
\end{figure}

%%%%%%%%%%%%%%%%%%%%%%%%%%%%%%%%%%%%%%%%%%%%%%%%%%%%%%%%%%%%%%%%%%%%%%%%%%%%
%%%%%%%%%%%%%%%%%%%%%%%%%%%%%%%%%%%%%%%%%%%%%%%%%%%%%%%%%%%%%%%%%%%%%%%%%%%%

\begin{table}[htbp]
	\caption{Compilation of DSC observations extracted from Fig.~\ref{fig:FIG1} using the same color code.}
	\label{tab:TAB2}
	\begin{ruledtabular}
		\begin{tabular}{l|c|c|c}
			Observation                                                           & temperature range {[}K{]} & $\Delta$H {[}J/g{]} & $\Delta$T$_g$ {[}K{]} \\ \hline
			\color{blue}{(1) exothermic peak} & $ 633 - 673 $                 & $ -1.34 $              & $ - $                     \\
			\color{blue}{(2) rejuvenation} & $ 120 - 300 $                 & $ -5.14 $              & $ - $                     \\ \hline
			\color{purple}{(3) exothermic peak} & $ 633 - 673 $                 & $ -0.49 $              & $ - $                     \\
			(4) shift in T$_g$ & $ \textbf{578 - \color{blue}{583} \color{black}{/} \color{red}{570} \color{black}{-} \color{purple}{576}} $     & $ - $                   & $ 5 / 6 $                 \\
			\color{purple}{(5) endothermic peak} & $ 618 - 628 $                 & $ +0.24 $              & $ - $                    
		\end{tabular}
	\end{ruledtabular}
\end{table} 

\subsection{Low-temperature heat capacity measurements (PPMS)}
Fig.~\ref{fig:FIG2} shows that deformation increased the boson peak (rejuvenation) compared with the as-cast state while annealing below the glass transition decreased the boson peak for both deformed and as-cast state (aging/relaxation) to the same level. However, their calorimetric responses are different (Fig.~\ref{fig:FIG1}b). Thus, the boson peak height (low-temperature excess of atomic vibration contributions to the Debye law) cannot explain the observed calorimetric differences in terms of re-distribution of excess free volume. In fact, different MRO types and relative volume fractions may explain these changes.

%%%%%%%%%%%%%%%%%%%%%%%%%%%%%%%%%%%%%%%%%%%%%%%%%%%%%%%%%%%%%%%%%%%%%%%%%%%%
%%%%%%%%%%%%%%%%%%%%%%%%%%%%%%%%%%%%%%%%%%%%%%%%%%%%%%%%%%%%%%%%%%%%%%%%%%%%

\subsection{Fluctuation Electron Microscopy}

\begin{figure*}[htbp]
	\includegraphics[width=\textwidth]{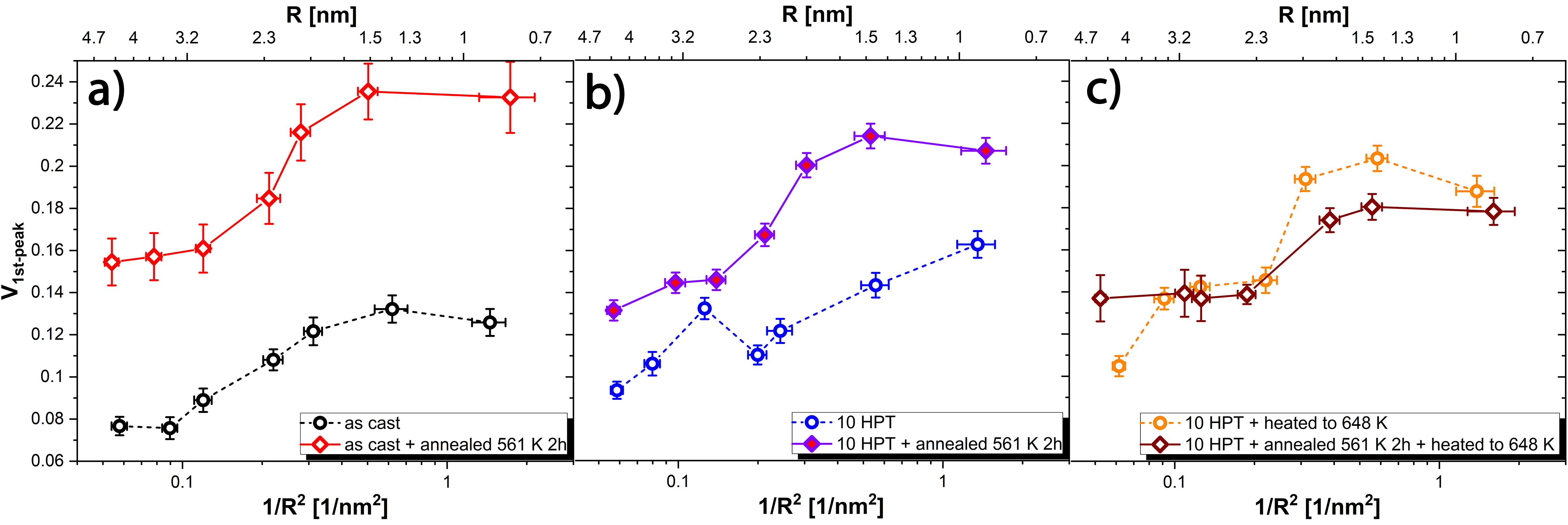}%
	\centering
	\caption{FEM analysis of the corresponding amorphous materials' states structure showing V$_{\mathrm{first}}$ at $k = 4.8$\,nm$^{-1}$ plotted against $1/R^2$: a) as-cast and annealed, b) 10 HPT and annealed and c) the ''frozen-in`` states of the exothermic peak.}
	\label{fig:FIG3}
\end{figure*}

\begin{figure}[htbp]
	\includegraphics[width=\columnwidth]{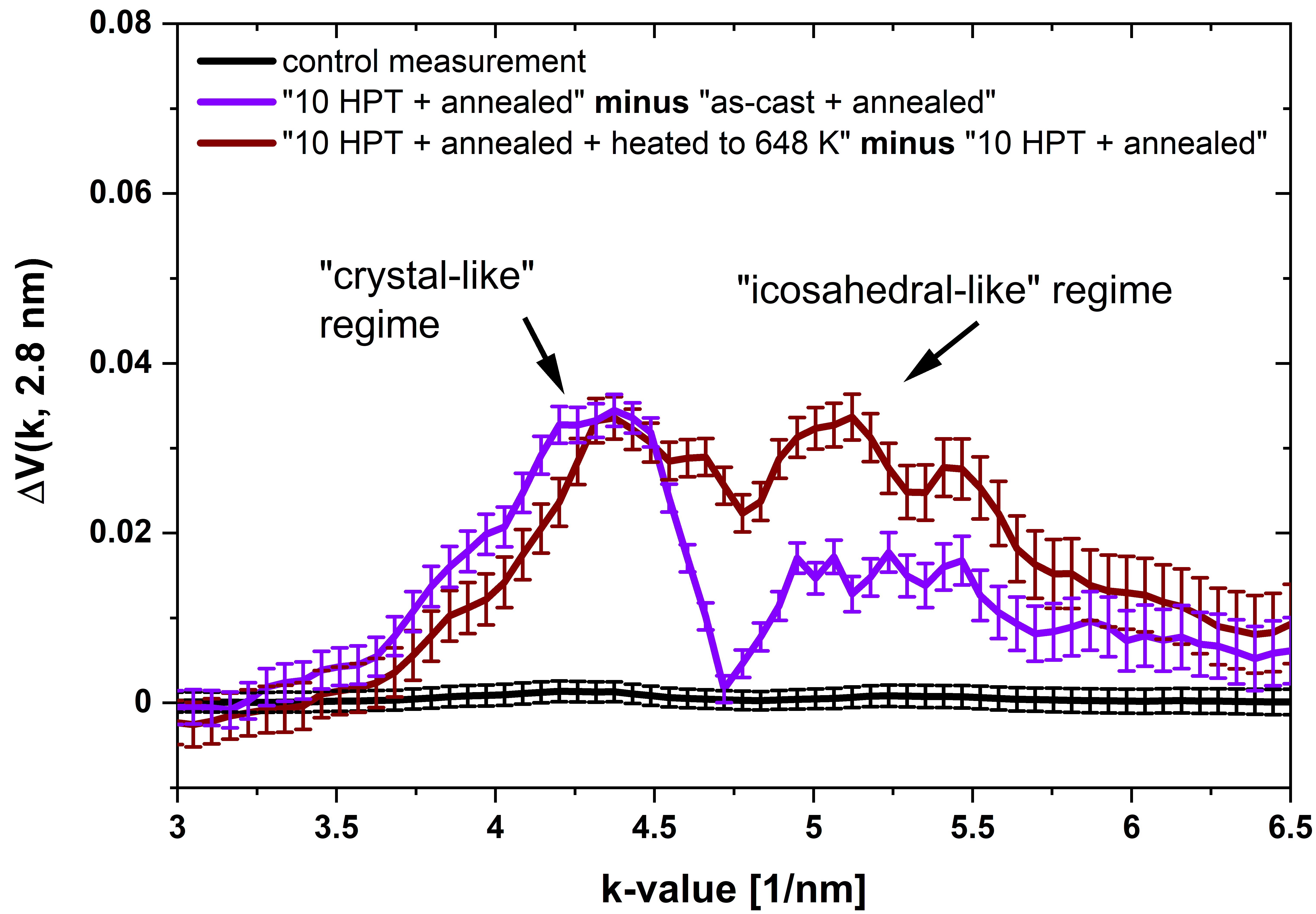}%
	\centering
	\caption{Differential curves of the normalized variance $\Delta$V(k, R$=2.8$\,nm) showing significant differences originating from the exothermic peaks. ''Icosahedral-`` and ''crystal-like`` dominated MRO types can be discerned.}
	\label{fig:FIG6}
\end{figure}

In Fig.~\ref{fig:FIG3} the normalized variance data V$_{\mathrm{first}}$ ($k_{\mathrm{1st-peak}} = 4.8$\,nm$^{-1}$) are plotted against $(1/R^2)$ \cite{voyles2000fluctuation,stratton2008phenomenological}. The obtained MRO results are presented as follows: 

\begin{itemize}
	\item[(i)] MRO sizes displayed in the form of plateaus
	\item[(ii)] relative MRO volume fractions estimated by the peak heights
	\item[(iii)] MRO type changes analyzed using the differential variance curves (Fig.~\ref{fig:FIG6}) for representative materials' states.
\end{itemize}

%%%%%%%%%%%%%%%%%%%%%%%%%%%%%%%%%%%%%%%%%%%%%%%%%%%%%%%%%%%%%%%%%%%%%%%%%%%%

Firstly, the ''as-cast`` state in Fig.~\ref{fig:FIG3}a reveals two ''plateaus`` in accordance with Refs. \cite{davani2020shear,hilke2020role,zhou2020x}. In comparison to the ''as-cast`` and ''as-cast + annealed 561\,K for 2\,h`` states (see Fig.~\ref{fig:FIG3}a), the ''10 HPT`` state shows no distinct plateau but rather a continuous decrease of the V$_{\mathrm{first}}$ signal with a peak at 2.8\,nm probe size (blue data points in Fig.~\ref{fig:FIG3}b). However, after annealing the ''10 HPT`` sample (purple data points in Fig.~\ref{fig:FIG3}b) the distinct peak for 2.8\,nm probe size disappears. Thus, the data reveal the existence of two MRO size distributions \cite{davani2020shear,hilke2020role,zhou2020x} with the larger MRO size more strongly affected by the imparted HPT deformation. Moreover, the two sample states which have been rapidly quenched from the exothermic peak (orange and brown data points in Fig.~\ref{fig:FIG3}c) show two plateaus overlapping to some extend, which is later discussed.

%%%%%%%%%%%%%%%%%%%%%%%%%%%%%%%%%%%%%%%%%%%%%%%%%%%%%%%%%%%%%%%%%%%%%%%%%%%%

Secondly, annealing raised the magnitude of V$_{\mathrm{first}}$ with respect to the non-annealed materials' states as shown in Fig.~\ref{fig:FIG3}a for as-cast (black and red data points) and Fig.~\ref{fig:FIG3}b for 10 HPT (blue and purple data points). Thus, the relative MRO volume fraction(s) increased by relaxation/aging \cite{zhang2016medium}. The magnitude of V$_{\mathrm{first}}$ ''10 HPT`` (blue data points in Fig.~\ref{fig:FIG3}b) with respect to the ''as-cast`` (black data points in Fig.~\ref{fig:FIG3}a) state has mainly increased within the plateaus, while a decrease in relative MRO volume fraction was observed for less severely deformed BMGs \cite{davani2020shear,hilke2019influence}. 

\noindent However, the two sample states which have been rapidly quenched from the exothermic peak (orange and brown data points in Fig.~\ref{fig:FIG3}c) show similar magnitudes and thus relative MRO volume fractions. Moreover, it is worth noting that the ''10 HPT + annealed 561\,K for 2\,h`` curve (purple data points in Fig.~\ref{fig:FIG3}b) and the ''10 HPT + heated to 648\,K`` (orange data points in Fig.~\ref{fig:FIG3}c) curve show great similarities (trend of plateaus and magnitude) with respect to MRO sizes and relative volume fractions indicating that the annealing procedure has a similar effect as heating into the supercooled liquid followed by rapid quenching. 

%%%%%%%%%%%%%%%%%%%%%%%%%%%%%%%%%%%%%%%%%%%%%%%%%%%%%%%%%%%%%%%%%%%%%%%%%%%%

Thirdly, the differential variance curves reveal structural changes in MRO types. Representative differential curves $\Delta V$ for the most dominantly affected probe size of 2.8\,nm are shown in Fig.~\ref{fig:FIG6}. The reproducibility and sensitivity of the differential signal is demonstrated in a control measurement using two independently recorded ''as-cast`` data sets (black curve in Fig.~\ref{fig:FIG6}).

\noindent Thus, the differential signal $\Delta$V(k, R$=2.8$\,nm) affords a discrimination into two different MRO types dominant at different k-values, i.e. $k = 5.3$\,nm$^{-1}$ - ''icosahedral-like`` and $k = 4.5$\,nm$^{-1}$ - ''crystal-like`` for NiP \cite{sheng2006atomic,wen2009distinguishing,zhan2017effect} and PdNiP \cite{wang2018spatial,hosokawa2019partial,du2019reentrant}. With respect to the ''crystal-like`` regime, it should be noted that the presence of nano-crystals is excluded since all NBDPs exhibit only speckle patterns \cite{treacy1996variable,treacy2007structural} for all six sample states, i.e. no distinct crystalline peaks were observed in any diffraction pattern. Thus, the observed differences originate only from modified dominant MRO types.

\noindent To elucidate the nature of the exothermic peak, the ''as-cast + annealed`` and ''10 HPT + annealed`` states were used as references from which the states displaying the exothermic peak in the supercooled liquid are subtracted. The corresponding two differential signals are shown in Fig.~\ref{fig:FIG6}. 

\noindent First, the purple curve in Fig.~\ref{fig:FIG6} represents the differential signal between the ''10 HPT + annealed`` and the ''as-cast + annealed`` state (see also the purple and red curves in Figs.~\ref{fig:FIG1}b and \ref{fig:FIG2}) prior to heating into the exothermic peak. A pronounced ''crystal-like`` regime is observed.

\noindent Second, heating into the exothermic peak of the ''10 HPT + annealed`` (see brown curve in Fig.~\ref{fig:FIG6}) shows both, an increase of the ''icosahedral-like`` and ''crystal-like`` order.

%%%%%%%%%%%%%%%%%%%%%%%%%%%%%%%%%%%%%%%%%%%%%%%%%%%%%%%%%%%%%%%%%%%%%%%%%%%%
%%%%%%%%%%%%%%%%%%%%%%%%%%%%%%%%%%%%%%%%%%%%%%%%%%%%%%%%%%%%%%%%%%%%%%%%%%%%
%%%%%%%%%%%%%%%%%%%%%%%%%%%%%%%%%%%%%%%%%%%%%%%%%%%%%%%%%%%%%%%%%%%%%%%%%%%%
%%%%%%%%%%%%%%%%%%%%%%%%%%%%%%%%%%%%%%%%%%%%%%%%%%%%%%%%%%%%%%%%%%%%%%%%%%%%
%%%%%%%%%%%%%%%%%%%%%%%%%%%%%%%%%%%%%%%%%%%%%%%%%%%%%%%%%%%%%%%%%%%%%%%%%%%%

\section{Discussion}
In the following the effects caused by thermo-mechanical treatments are discussed with respect to the structural features obtained by FEM. Finally, the results are summarized in a PEL scheme. 

\subsection{Calorimetric observations by DSC and PPMS}
HPT deformation performed at room temperature affected the calorimetric signal significantly. Since the sensitivity of calorimetric measurements to phase transformations is in the order of several percent of the sample volume, it is concluded that a relevant volume fraction of the material has been changed to contribute to exothermic peaks, (1) and (3), in the supercooled liquid shown in Fig.~\ref{fig:FIG1}a,b.

The increase of the exothermic peak below T$_g$ is typically called rejuvenation \cite{ketov2015rejuvenation} (see observation (2) in Fig.~\ref{fig:FIG1}a) and is related to an increase of the excess free volume caused by severe plastic deformation \cite{zhou2019two,hubek2018impact}. Here the observed structural enthalpy change displaying the rejuvenation in Tab.~\ref{tab:TAB2} is in good agreement with a former report \cite{zhou2019two}. Moreover, the shift in T$_g$, here defined via the half-c$_p$ method \cite{hubek2018impact}, is an indicator that the individual relaxation state is influenced by both deformation ($\Delta$T$_g = 5 / 6$\,K - see black and blue as well as red and purple curves in Fig.~\ref{fig:FIG1}) and annealing (aging: $\Delta$T$_g = 7 / 8$\,K - see black and red as well as blue and purple curves in Fig.~\ref{fig:FIG1}).

%%%%%%%%%%%%%%%%%%%%%%%%%%%%%%%%%%%%%%%%%%%%%%%%%%%%%%%%%%%%%%%%%%%%%%%%%%%%

The data shown in Fig.~\ref{fig:FIG1}b clearly reveal that subsequent annealing below T$_g$ equalizes the relaxation signal. However, changes (observation (3) and (5)) in the supercooled liquid are observed. Their origin is now discussed.

\noindent Recent molecular dynamics (MD) simulations showed that at the glass transition the motif distribution in metal-metal MGs changed towards an increase in ''icosahedral-like`` order in the supercooled liquid \cite{maldonis2019short}. On the other hand, for metal-metalloid MGs the influence of covalent bonding seems to be the key parameter leading to a frustration of icosahedral structures \cite{maldonis2019local,zhan2017effect,moitzi2020chemical,Maldonis2019,Davani2019}. This different population of motifs \cite{maldonis2019local} may influence the thermal stability of certain MRO types in the BMG, even above T$_g$. Here, our observations indicate that strain-induced modifications had occurred, contributing significantly to the thermal stability of the glassy material.

\noindent Several studies have reported the existence of an exothermic peak occurring in the supercooled liquid which was related to a polymorphic transition between two liquids having different densities \cite{kim2013phase,sheng2007polyamorphism,ha1996supercooled}. This characteristic may also be classified a sort of polyamorphism \cite{sheng2007polyamorphism,zhu2015possible}. Kumar et al. \cite{kumar2006structural} performed TEM experiments on amorphous Zr$_{36}$Ti$_{24}$Be$_{40}$ suggesting that the occurrence of the exothermic peak is a consequence of short-range order rearrangements paving the way for crystallization. Novel findings by Lou et al. \cite{lou2020two} using in-situ synchrotron X-ray techniques also report an exothermic peak in the supercooled liquid. However, the observation of the exothermic peak was related to MRO changes involving ordering towards pre-nucleation sites \cite{lou2020two}.

\noindent Moreover, additional transformations in the supercooled liquid region of Pd$_{41.25}$Ni$_{41.25}$P$_{17.5}$ were observed by Lan et al. \cite{lan2017hidden} using in-situ X-ray diffraction. They attributed the presence of the exothermic peak in the supercooled liquid to changes in the phosphorous content finally influencing the MRO structure, while for the stoichiometric ternary Pd$_{40}$Ni$_{40}$P$_{20}$ model alloy this was not observed \cite{madge2005transformations}. 

\noindent In fact, the present study clearly revealed an exothermic peak in the supercooled liquid occurring after HPT deformation for the stoichiometric ternary Pd$_{40}$Ni$_{40}$P$_{20}$ model alloy. Hence, following the line of argumentation from the literature discussed above, the origin of this exothermic peak is most likely related to strain-induced modifications which exhibit a higher thermal stability. This interpretation is discussed with respect to structural aspects in more depth in the FEM section.

%%%%%%%%%%%%%%%%%%%%%%%%%%%%%%%%%%%%%%%%%%%%%%%%%%%%%%%%%%%%%%%%%%%%%%%%%%%%

Now we briefly discuss the endothermic signal appearing in the supercooled liquid after subsequent annealing, which may resemble a second glass transition (see observation (5) in Fig.~\ref{fig:FIG1}b). A second glass transition accompanied by an exothermic peak in the supercooled liquid was reported for Pd$_{42.5}$Ni$_{42.5}$P$_{15}$ \cite{du2019reentrant} and has been related to MRO ordering phenomena. It can be hypothesized that the observation in the present study may also be an ordering phenomenon, yet of extremely thermally stable MRO \cite{nollmann2016impact,hubek2018impact,hubek2020intrinsic,hilke2020role}.

%%%%%%%%%%%%%%%%%%%%%%%%%%%%%%%%%%%%%%%%%%%%%%%%%%%%%%%%%%%%%%%%%%%%%%%%%%%%

The calorimetric results obtained at cryogenic temperatures via PPMS measurements showed an enhancement of the boson peak after HPT deformation compared to the ''as-cast`` state (see Fig.~\ref{fig:FIG2}). These results are related to changes in the amount of excess free volume which is in agreement with the rejuvenation signal observed in the DSC data (see observation (2) in Fig.~\ref{fig:FIG1}a and Tab.~\ref{tab:TAB2}). Thus, the structure of the glassy material was modified by severe plastic deformation leading to an increased heterogeneity \cite{bernal1960geometry,elliott1995extended,torquato2000random,miracle2003influence,miracle2007structural}, in which different arrangements of atoms or clusters in space locally produce an increased amount of excess free volume \cite{mitrofanov2015impact,bunz2014low}. However, structural aging/relaxation generally leads to a decrease in the boson peak height \cite{hubek2018impact,zhou2019two}.

\noindent Strikingly, the present PPMS results revealed congruent boson peak curves after annealing the ''10 HPT`` and ''as-cast`` states (see Fig.~\ref{fig:FIG2}). However, the calorimetric responses in Fig.~\ref{fig:FIG1}b are different above the annealing temperature. As a consequence, the differences need to originate from different atomic arrangements rather than the absolute amount of excess free volume, defining the structural heterogeneity \cite{Davani2019,hilke2020role}. This motivates a detailed investigation of the amorphous structure, here characterized with respect to the MRO, in order to relate structural changes to the calorimetric signals.

%%%%%%%%%%%%%%%%%%%%%%%%%%%%%%%%%%%%%%%%%%%%%%%%%%%%%%%%%%%%%%%%%%%%%%%%%%%%
%%%%%%%%%%%%%%%%%%%%%%%%%%%%%%%%%%%%%%%%%%%%%%%%%%%%%%%%%%%%%%%%%%%%%%%%%%%%

\subsection{Fluctuation Electron Microscopy}
In this section the individual thermo-mechanical treatments affecting the MRO microstructure (sizes, volume fractions in Fig.~\ref{fig:FIG3} and types in Fig.~\ref{fig:FIG6}) are discussed with respect to the evolution of the exothermic peak.

%%%%%%%%%%%%%%%%%%%%%%%%%%%%%%%%%%%%%%%%%%%%%%%%%%%%%%%%%%%%%%%%%%%%%%%%%%%%

Firstly, the observation of two plateaus for Pd$_{40}$Ni$_{40}$P$_{20}$ shown in Fig.~\ref{fig:FIG3} is in very good agreement with former reports \cite{Davani2019,hilke2020role,zhou2020x}. While non-severe plastic deformation characterized by the formation of individual shear bands \cite{davani2020shear,rosner2014density,hilke2019influence,hieronymus2017shear,schmidt2015quantitative,hilke2020role} also showed two plateaus \cite{Davani2019}, the impact of HPT deformation appears in the form of a continuous decrease in MRO except the peak at 2.8\,nm probe size. This deviation can be explained by a very narrow MRO size distribution due to the severe deformation mode.

\noindent Moreover, the similarity of the purple curve in Fig.~\ref{fig:FIG3}b and the red one in Fig.~\ref{fig:FIG3}a fits to the observation of the congruent curves of the boson peaks shown in Fig.~\ref{fig:FIG2} confirming that annealing leads to similar MRO sizes (plateaus) and volume fractions estimated by the V$_\mathrm{first}$ magnitudes. Albeit, as the corresponding calorimetric signals are different in the supercooled liquid (Fig.~\ref{fig:FIG1}b), it is concluded that the MRO sizes alone cannot explain the observed differences.

%%%%%%%%%%%%%%%%%%%%%%%%%%%%%%%%%%%%%%%%%%%%%%%%%%%%%%%%%%%%%%%%%%%%%%%%%%%%

Secondly, annealing (aging/relaxation) increases the relative MRO volume fractions which is in good agreement with previously reported results \cite{bogle2007quantifying,bogle2010size,zhang2016medium}. However, non-severe deformation led to a decrease in the relative MRO volume fractions for shear band environments \cite{hilke2019influence,davani2020shear} which appears, at first glance, to be contrary to the present results. A plausible explanation may be found in the severe deformation mode which leads to very complex materials' states. As discussed above, this also fits to the interpretation of very narrow MRO size distributions. 

\noindent Interestingly, both rapidly quenched-in materials' states featuring the exothermic peak in Fig.~\ref{fig:FIG3}c show a similar relative MRO volume fraction. These seem to be correlated since both states were quenched with identical cooling rates. An upper boundary for the relative MRO volume fractions appears reasonable with respect to the annealed states shown in Fig.~\ref{fig:FIG3}a,b which may also represent a highly relaxed state. Moreover, this agrees with the observation that the ''10 HPT + annealed at 561\,K for 2\,h`` state (see purple chart in Fig.~\ref{fig:FIG3}b) leads to similar MRO sizes and relative volume fractions as the ''10 HPT + heated to 648\,K`` state (see orange chart in Fig.~\ref{fig:FIG3}c). Since the exothermic peak in the supercooled liquid has been discussed above with respect to the complex interplay of thermo-mechanical treatments with the MRO sizes and relative volume fractions, a further discrimination in terms of MRO types is made. 

%%%%%%%%%%%%%%%%%%%%%%%%%%%%%%%%%%%%%%%%%%%%%%%%%%%%%%%%%%%%%%%%%%%%%%%%%%%%

\begin{figure*}[htbp]
	\includegraphics[width=\textwidth]{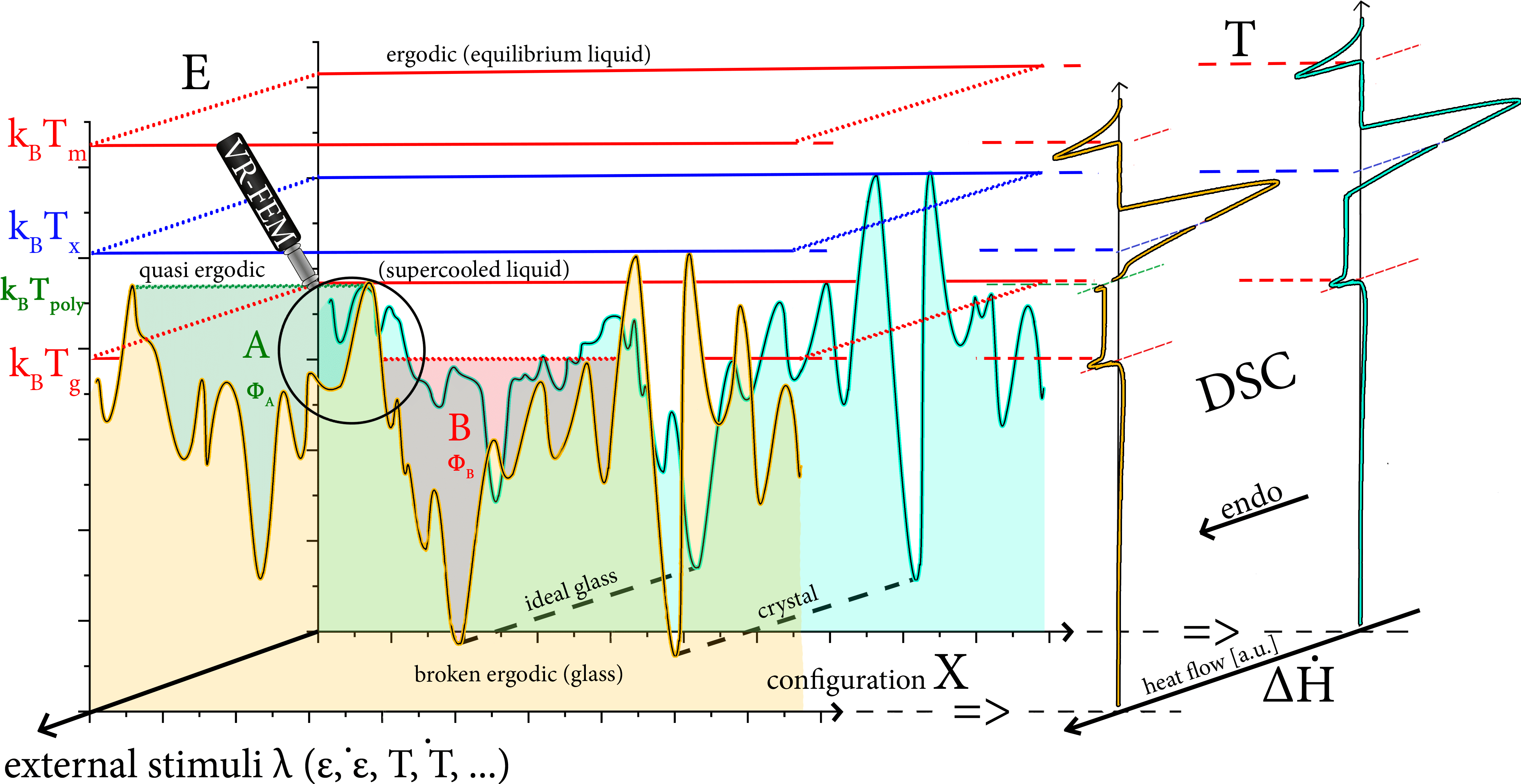}%
	\centering
	\caption{A schematic 2D illustration of PELs representing the ''as-cast`` state and the ''10 HPT deformed + annealed`` state. The dependence of the total potential energy $ E $ is shown as a function of the configurational parameter $X $ and external stimuli $\lambda(\epsilon, \dot{\epsilon}, T, \dot{T}, ...)$ corresponding to the thermo-mechanical history. Region A (green meta-basin) displays the higher thermally stable ''crystal-like`` MRO network. Region B (red meta-basin) indicates the glass transition T$_g$ for the ''icosahedral-like`` MRO network.}
	\label{fig:FIG4}
\end{figure*}

Thirdly, trends in the MRO types are exemplarily shown by the differential signal $\Delta$V(k, R$=2.8$\,nm) in Fig.~\ref{fig:FIG6} and are now discussed regarding the occurrence of the exothermic peak in the supercooled liquid. Since the previous discussion pointed out that the exothermic peak only occurs after severe plastic deformation, changes in the MRO type are expected. Two dominant MRO types at $k = 5.3$\,nm$^{-1}$ as ''icosahedral-like`` and at $k = 4.5$\,nm$^{-1}$ as ''crystal-like`` were identified \cite{sheng2006atomic,wen2009distinguishing,zhan2017effect,wang2018spatial,hosokawa2019partial,du2019reentrant}. Due to either deformation or annealing the population between ''crystal-`` and ''icosahedral-like`` order was changed. By sub-T$_g$ annealing of the HPT deformed state (see purple curve in Fig.~\ref{fig:FIG6}) a significant increase in ''crystal-like`` order was achieved. It is emphasized, although the red and purple curves appear similar in the DSC and PPMS measurements shown in Fig.~\ref{fig:FIG1}b and  Fig.~\ref{fig:FIG2} until the annealing temperature of 561\,K, the MRO types are significantly changed in Fig.~\ref{fig:FIG6}. Moreover, while the ''icosahedral-like`` order increases when going through the glass transition \cite{maldonis2019short,maldonis2019local} (heated to 648 \,K), the ''crystal-like`` order was as well increased. Thus, the difference of the DSC signal in the supercooled liquid region in Fig.~\ref{fig:FIG1}b is most likely related to an increase of ''crystal-like`` MRO.

%%%%%%%%%%%%%%%%%%%%%%%%%%%%%%%%%%%%%%%%%%%%%%%%%%%%%%%%%%%%%%%%%%%%%%%%%%%%

To summarize our observations, the occurrence of the exothermic peak in the supercooled liquid is caused by severe plastic deformation and is related to MRO type modifications. On the other hand, pure annealing treatments lead only to changes in the relative MRO volume fractions and sizes. Moreover, the combination of calorimetric and structural observations suggest that ''crystal-like`` MRO represent a higher thermally stable state which may act potentially as embryos or seeds for nucleation. This indicates the presence of two different MRO networks \cite{lee2011networked} composed of ''icosahedral`` and ''crystal-like`` regimes. 

\noindent The PEL of a BMG represented by a complex hierarchy of local minima is thus well-suited to explain our observations assuming here the individual minima of the PEL to be associated with certain MRO networks (types). 

%%%%%%%%%%%%%%%%%%%%%%%%%%%%%%%%%%%%%%%%%%%%%%%%%%%%%%%%%%%%%%%%%%%%%%%%%%%%
%%%%%%%%%%%%%%%%%%%%%%%%%%%%%%%%%%%%%%%%%%%%%%%%%%%%%%%%%%%%%%%%%%%%%%%%%%%%

\subsection{Results in the context of a Potential Energy Landscape (PEL)}
The relaxation and structural modifications in BMGs caused by either deformation or annealing are described in terms of local and semi-global minima, called ''basins`` and ''meta-basins``, typically illustrated in two-dimensional PELs  \cite{malandro1999relationships,xu2018effects,kuchemann2018energy,stillinger1982hidden,heuer2008exploring}.

\noindent Annealing provides the necessary thermal energy for the system to surmount local energy barriers and thus transform from one local minimum into a deeper one within a meta-basin. 

\noindent In contrast, deformation, as a directional way of increasing the potential energy by strain, biases the entire PEL by warping the trajectories in 3(N-1) dimensions. Here N refers to the number of atoms. As a consequence, the energy barriers are significantly lowered under the external stimulus affording transitions between individual meta-basins. This is now called ''inter meta-basins crossings``. Such minima exhibit a higher thermal stability \cite{stillinger1982hidden,heuer2008exploring} as observed here for the severe plastic deformed and subsequently annealed BMG. 

%%%%%%%%%%%%%%%%%%%%%%%%%%%%%%%%%%%%%%%%%%%%%%%%%%%%%%%%%%%%%%%%%%%%%%%%%%%%

In this study, a combination of both deformation and subsequent annealing was applied resulting in the occurrence of an exothermic peak. The corresponding PEL can be probed by integral calorimetric methods such as DSC or PPMS and are shown as integral projections on the right-hand side in Fig.~\ref{fig:FIG4}. Generally, DSC or PPMS quantifies always an averaged signal over the whole volume. 

However, FEM is sensitive to local structural features in terms of MRO and its changes caused by deformation or annealing. Using FEM the individual ''icosahedral-/crystal-like`` MRO networks (types) were identified for the different sample states. As a consequence, firstly, the ''crystal-like`` MRO ordering is attributed to the exothermic peak and secondly, the two MRO types are necessarily linked to different meta-basins. This is illustrated within the framework of PELs by incorporating the corresponding external stimuli ($\lambda(\epsilon, \dot{\epsilon}, T, \dot{T}, ...)$) shown in Fig.~\ref{fig:FIG4}. 

The initial ''as-cast`` state is illustrated by the turquoise background in Fig.~\ref{fig:FIG4}. The material exhibits here a much larger volume fraction of configuration \textbf{B} (''icosahedral-like`` MRO network) characterized by the MRO volume fraction $\Phi_B$ than \textbf{A} (''crystal-like`` MRO network) characterized by the MRO volume fraction $\Phi_A$. Deformation warps the  PEL such that a fraction of the ''icosahedral-like`` MRO network reaches a higher thermally stable meta-basin. This can also be interpreted as rejuvenation extending the current calorimetric definition. Yet, for Pd$_{40}$Ni$_{40}$P$_{20}$ rejuvenation by cryogenic thermal cycling was not observed \cite{Nollmann2018}. In Fig.~\ref{fig:FIG4} region \textbf{A} represents a stable meta-basin (higher kinetic stability) within the glass transforming at T$_{poly} = 648$\,K in the form of the exothermic peak.

%%%%%%%%%%%%%%%%%%%%%%%%%%%%%%%%%%%%%%%%%%%%%%%%%%%%%%%%%%%%%%%%%%%%%%%%%%%%

\noindent Moreover, region \textbf{A} in Fig.~\ref{fig:FIG4} may also resemble a first-order transition of two amorphous phases (polyamorphism) \cite{zhu2015possible, Maldonis2019, lou2020two}, represented in our case by the two different MRO networks. However, our data does not reveal a reentrant enthalpy change towards an ultra-stable second supercooled liquid configuration as reported for off-stoichiometric Pd$_{41.25}$Ni$_{41.25}$P$_{17.5}$ \cite{du2019reentrant}. Our interpretation regarding a first-order transition is supported by a recent ab-initio MD simulations on composition-dependent investigations of the Pd-Ni-P system \cite{wang2020anomalous} as well as by a Reverse Monte Carlo study \cite{du2019reentrant} claiming a MRO ordering towards pre-nucleation sites in the form of ''crystal-like`` embryos. 

%%%%%%%%%%%%%%%%%%%%%%%%%%%%%%%%%%%%%%%%%%%%%%%%%%%%%%%%%%%%%%%%%%%%%%%%%%%%
%%%%%%%%%%%%%%%%%%%%%%%%%%%%%%%%%%%%%%%%%%%%%%%%%%%%%%%%%%%%%%%%%%%%%%%%%%%%
%%%%%%%%%%%%%%%%%%%%%%%%%%%%%%%%%%%%%%%%%%%%%%%%%%%%%%%%%%%%%%%%%%%%%%%%%%%%
%%%%%%%%%%%%%%%%%%%%%%%%%%%%%%%%%%%%%%%%%%%%%%%%%%%%%%%%%%%%%%%%%%%%%%%%%%%%

\section{Conclusions}
This work investigated the impact of severe plastic deformation by high-pressure torsion on the relaxation of the glassy and supercooled liquid states of Pd$_{40}$Ni$_{40}$P$_{20}$ using DSC, low-temperature heat capacity measurements and FEM. The changes in the calorimetric signals due to deformation and subsequent heat treatments were analyzed and a correlation between deformation (rejuvenation) and annealing (relaxation) was found in relation to MRO. An exothermic peak was observed in the supercooled liquid region which could be attributed to changes in the MRO types. These findings are comprehended in a potential energy landscape scheme offering a new approach for MRO engineering of glasses.

%%%%%%%%%%%%%%%%%%%%%%%%%%%%%%%%%%%%%%%%%%%%%%%%%%%%%%%%%%%%%%%%%%%%%%%%%%%%
%%%%%%%%%%%%%%%%%%%%%%%%%%%%%%%%%%%%%%%%%%%%%%%%%%%%%%%%%%%%%%%%%%%%%%%%%%%%
%%%%%%%%%%%%%%%%%%%%%%%%%%%%%%%%%%%%%%%%%%%%%%%%%%%%%%%%%%%%%%%%%%%%%%%%%%%%
%%%%%%%%%%%%%%%%%%%%%%%%%%%%%%%%%%%%%%%%%%%%%%%%%%%%%%%%%%%%%%%%%%%%%%%%%%%%

\section*{SUPPLEMENTARY MATERIAL}
See Supplementary Material for the extended data sets of the variable resolution fluctuation electron microscopy acquisition.

\section*{AUTHOR'S CONTRIBUTIONS}
All authors contributed equally to this work.

\section*{DATA AVAILABILITY}
The original data of this study are available from the corresponding author upon reasonable request. Moreover, the analyzed data that supports the findings of this study are available within the article and its Supplementary Material.

%%%%%%%%%%%%%%%%%%%%%%%%%%%%%%%%%%%%%%%%%%%%%%%%%%%%%%%%%%%%%%%%%%%%%%%%%%%%
%%%%%%%%%%%%%%%%%%%%%%%%%%%%%%%%%%%%%%%%%%%%%%%%%%%%%%%%%%%%%%%%%%%%%%%%%%%%
%%%%%%%%%%%%%%%%%%%%%%%%%%%%%%%%%%%%%%%%%%%%%%%%%%%%%%%%%%%%%%%%%%%%%%%%%%%%
%%%%%%%%%%%%%%%%%%%%%%%%%%%%%%%%%%%%%%%%%%%%%%%%%%%%%%%%%%%%%%%%%%%%%%%%%%%%

\begin{acknowledgments}
We gratefully acknowledge financial support by the DFG via WI 1899/29-1. The DFG is further acknowledged for funding our TEM equipment via the Major Research Instrumentation Program under INST 211/719-1 FUGG. 
\end{acknowledgments}

\section*{REFERENCES}
\bibliography{literature}

\end{document}